\documentclass[twocolumn,pra,showpacs,aps,superscriptaddress]{revtex4}

\usepackage{amsmath}
\usepackage{amsbsy}
\usepackage{amsthm}
\usepackage{amssymb}
\usepackage{latexsym}
\usepackage{textcomp}
\usepackage[dvips]{color,graphicx}

\newcommand{\expv}[1]{\langle #1 \rangle}	
\newcommand{\ket}[1]{| #1 \rangle} 
\newcommand{\bra}[1]{\langle #1 |} 
\newcommand{\op}[1]{\hat #1}

\begin{document}

\title{Initial state dependence of the quench dynamics in integrable quantum systems}

\author{Marcos Rigol}
\affiliation{Department of Physics, Georgetown University, Washington, DC 20057, USA}
\author{Mattias Fitzpatrick}
\affiliation{Department of Physics, Georgetown University, Washington, DC 20057, USA}
\affiliation{Department of Physics, Middlebury College, Middlebury, Vermont 05753, USA}

\begin{abstract}
We identify and study classes of initial states in integrable quantum systems that,
after the relaxation dynamics following a sudden quench, lead to near-thermal expectation 
values of few-body observables. In the systems considered here, those states are found to be 
insulating ground states of lattice hard-core boson Hamiltonians. We show that, as a suitable 
parameter in the initial Hamiltonian is changed, those states become closer to Fock states 
(products of single site states) as the outcome of the relaxation dynamics becomes closer 
to the thermal prediction. At the same time, the energy density approaches a Gaussian. 
Furthermore, the entropy associated with the generalized canonical and generalized grand-canonical 
ensembles, introduced to describe observables in integrable systems after relaxation, approaches 
that of the conventional canonical and grand-canonical ensembles. We argue that those classes 
of initial states are special because a control parameter allows one to tune the distribution of 
conserved quantities to approach the one in thermal equilibrium. This helps in understanding the 
approach of all the quantities studied to their thermal expectation values. However, a finite-size 
scaling analysis shows that this behavior should not be confused with 
thermalization as understood for nonintegrable systems.
\end{abstract}
\maketitle

\section{Introduction}

The relaxation dynamics of isolated quantum systems after a sudden quench is a topic that is 
attracting much current attention. Interest on this problem has been sparked by recent 
experiments with ultracold gases 
\cite{greiner_mandel_02b,kinoshita_wenger_06,hofferberth_lesanovsky_07,trotzky_chen_11}.
The high degree of isolation in those experiments allows one to consider them as almost ideal 
analog simulators of the unitary dynamics of pure quantum states. For example, in 
Ref.~\cite{kinoshita_wenger_06}, Kinoshita {\it et al.}~showed that observables in a 
(quasi-)one-dimensional bosonic system close to an integrable point do not relax to the 
values expected from a conventional statistical mechanics description. Any non-negligible 
coupling to a thermal environment would have destroyed such a remarkable phenomenon. More 
recently, Trotzky {\it et al.}~\cite{trotzky_chen_11} have shown that the experimental 
dynamics of Bose-Hubbard like (quasi-)one-dimensional systems can be almost perfectly 
described by the unitary dynamics of the relevant model Hamiltonian. The latter was 
followed by numerically exact means utilizing the time-dependent renormalization group 
algorithm \cite{white_feiguin_04,schollwock_review_05}.  

After the experimental results in Ref.~\cite{kinoshita_wenger_06}, many theoretical works
have found that, following a sudden quench within integrable systems, few-body observables, 
in general, relax to nonthermal expectation values
\cite{rigol_dunjko_07,rigol_muramatsu_06,cazalilla_06,calabrese_cardy_07a,cramer_dawson_08,
barthel_schollwock_08,eckstein_kollar_08,kollar_eckstein_08,rossini_silva_09,iucci_cazalilla_09,
fioretto_mussardo_10,iucci_cazalilla_10,mossel_caux_10,rossini_susuki_10,cassidy_clark_11,
calabrese_essler_11,cazalilla_iucci_11} (for a recent review, see Ref.~\cite{polkovnikov_sengupta_11}). 
Some of the novel insights gained through these studies 
include (i) the possibility of describing observables after relaxation by means of generalized
Gibb ensembles (GGE) \cite{rigol_dunjko_07,rigol_muramatsu_06,cazalilla_06,calabrese_cardy_07a,
cramer_dawson_08,barthel_schollwock_08,eckstein_kollar_08,kollar_eckstein_08,iucci_cazalilla_09,
fioretto_mussardo_10,iucci_cazalilla_10,cassidy_clark_11, calabrese_essler_11,
cazalilla_iucci_11}; (ii) the fact that even though in some 
cases the behavior of nonlocal observables after relaxation can be parametrized similarly to 
the one in thermal equilibrium \cite{rossini_silva_09,rossini_susuki_10}, an exact description of those 
observables is provided only by the GGE \cite{calabrese_essler_11},
and (iii) an understanding of the GGE through a generalization of the eigenstate thermalization
hypothesis (ETH) \cite{cassidy_clark_11}. ETH explains why thermalization occurs in generic 
(nonintegrable) quantum systems after a quench \cite{deutsch_91,srednicki_94,rigol_dunjko_08}.

All the results discussed above have been obtained in studies of several specific models. 
However, they are expected to hold in general for integrable systems. An interesting, 
and so far nongeneric, result reported in Refs.~\cite{rigol_muramatsu_06,cassidy_clark_11}
was the observation of a phenomenon close to ``real'' thermalization in integrable systems, 
in the sense of the expectation values of few-body observables after relaxation approaching those 
predicted in thermal equilibrium. This occurred as a parameter used to generate special classes 
of initial states was changed. In Ref.~\cite{rigol_muramatsu_06}, the initial states 
were insulating ground states of hard-core bosons in half-filled period-two superlattices, 
while in Ref.~\cite{cassidy_clark_11}, they were the ground state of trapped systems with a 
Mott insulating domain in the trap center.

In this work, we revisit the systems above and focus on understanding the properties 
of the initial states for which observables after relaxation were seen to approach 
thermal expectation values, despite integrability. As said before, those states are insulating 
ground states. Here, we show that the selected tuning parameter makes those initial states 
approach Fock states (products of single site wavefunctions) at the same time that (i) their 
energy density approaches a Gaussian, and (ii) the entropy of their associated generalized 
canonical and grand-canonical ensembles approach the entropies of the canonical and 
grand-canonical ensembles. We argue that (i) and (ii) above can be understood because the 
distribution of conserved quantities in such initial states approaches the one of systems 
in thermal equilibrium. Hence, they can have thermal-like energy densities, entropies, 
and observables after relaxation. However, after a finite-size scaling analysis, we conclude that 
this phenomenon differs conceptually from thermalization as it happens in 
nointegrable systems.

The presentation is organized as follows: in Sec.~\ref{sec:models}, we introduce the 
models and observables of interest. We also define the ensembles considered 
and provide details on how the calculations are performed. Section \ref{sec:overlaps} is devoted to 
study of the overlaps of the initial states with the eigenstates of the final 
Hamiltonians, as well as to the description of the energy densities in all cases. 
The scaling of the entropy with system size, for the different ensembles analyzed and for 
superlattice and trapped systems, is presented in Sec.~\ref{sec:entropies}.
In Sec.~\ref{sec:conserved}, we study the distribution of the conserved quantities for the 
different initial states and within standard statistical ensembles. Finally, the 
conclusions are presented in Sec.~\ref{sec:summary}.

\section{Model, ensembles, and observables} \label{sec:models}

We are interested in the equilibrium and nonequilibrium 
properties of lattice bosons in the limit of infinite on-site repulsion 
(hard-core bosons). Those systems can be described by the Hamiltonian
\begin{equation}
\hat{H}= - J \sum_{j=1}^{L-1} 
\left( \hat{b}_{j}^{\dagger} \hat{b}_{j+1} + \textrm{H.c.} \right)
 + \sum_{j=1}^{L}V^{\textrm{ext}}_{j} \hat{n}_{j},
\label{eq:hamilt}
\end{equation}
with the additional on-site constraints $\hat{b}_{j}^{\dagger2}=\hat{b}_{j}^{2}=0$,
which preclude multiple occupancy of the lattice sites. Here, $J$ is the 
nearest-neighbor hopping, $V^{\textrm{ext}}_j$ is a site-dependent local potential, and $L$
is the number of lattice sites. The hard-core 
boson creation (annihilation) operator in each site is denoted by 
$\hat{b}^{\dagger}_{j}$ ($\hat{b}_{j}$) and the site number occupation by 
$\hat{n}_{j}=\hat{b}^{\dagger}_{j} \hat{b}_{j}$. In what follows, we consider only 
systems with open boundary conditions, and $t=1$ sets our units of energy. 

This model is integrable \cite{lieb_shultz_61} and can be exactly solved by first 
mapping it onto the spin-1/2 $XX$ model (with a site-dependent magnetic field in the
$z$ direction) by means of the Holstein-Primakoff transformation
\cite{holstein_primakoff_40} and then onto a noninteracting fermion Hamiltonian
utilizing the Jordan-Wigner transformation \cite{jordan_wigner_28,lieb_shultz_61}.
In the fermionic language, the Hamiltonian can be straightforwardly diagonalized and 
all the spectral and thermodynamic properties of hard-core bosons can be computed 
either analytically or numerically in polynomial time. Off-diagonal correlations
are more difficult to calculate. However, using properties of Slater determinants, 
they can also be computed very efficiently numerically for ground state 
\cite{rigol_muramatsu_05HCBb,he_rigol_11} and finite temperature \cite{rigol_05} 
equilibrium problems, as well as during the 
unitary nonequilibrium dynamics \cite{rigol_muramatsu_05eHCBc}. Those 
insights will be used later.

More generally, the nonequilibrium dynamics of isolated quantum systems can be studied by 
writing the (arbitrary) initial state $\ket{\psi_I}$ as a linear combination of the eigenstates 
$\ket{\Psi_\alpha}$ of the Hamiltonian $\op{H}$ that drives the dynamics, 
which satisfies $\op{H}\ket{\Psi_\alpha}=E_\alpha\ket{\Psi_\alpha}$. Hence,
\begin{equation}
\ket{\psi_I}=\sum_{\alpha=1}^D C_\alpha \ket{\Psi_\alpha},
\label{eq:inistate}
\end{equation}
where $D$ is the 
dimension of the Hilbert space and $C_\alpha=\langle \Psi_\alpha \ket{\psi_I}$, 
and the time evolving wave function can be 
written as 
\begin{equation}
\ket{\Psi(t)}=e^{-i \op{H}t/\hbar}\ket{\psi_I}=
\sum_{\alpha=1}^D C_{\alpha}e^{-iE_\alpha t/\hbar} \ket{\Psi_\alpha}.
\label{eq:wf}
\end{equation}

The time evolution of a generic observable $\op{O}$ is then dictated by the 
sums over all eigenstates
\begin{equation}
\expv{\op{O}(t)} = \bra{\Psi(t)}\op{O}\ket{\Psi(t)}
=\sum_{\alpha,\beta} C^*_\alpha C_\beta\, e^{i(E_\alpha-E_\beta) t/\hbar}\, O_{\alpha\beta},
\label{eq:dynam}
\end{equation}
where $O_{\alpha\beta}=\bra{\Psi_\alpha}\op{O}\ket{\Psi_\beta}$, and the infinite 
time average of Eq.~\eqref{eq:dynam} (in the absence of degeneracies) can be thought as the result 
of a diagonal ensemble average \cite{rigol_dunjko_08}
\begin{equation}
\expv{\hat{O}}_\text{DE} =\sum_\alpha |C_\alpha|^2 O_{\alpha\alpha}.
\label{eq:diag}
\end{equation}
As shown in Refs.~\cite{rigol_dunjko_08,cassidy_clark_11}, this infinite time average 
describes observables after relaxation. We should stress that this can be true even in 
the presence of degeneracies associated with integrability, except for cases with massive 
degeneracies \cite{kollar_eckstein_08}. The validity of the description of integrable systems 
after relaxation, by means of the infinite time average \eqref{eq:diag}, has been demonstrated 
for the $1/r$ Hubbard model in Ref.~\cite{kollar_eckstein_08}, and for the same (hard-core boson) 
systems considered here in Ref.~\cite{cassidy_clark_11} (supplementary materials).

The result in Eq.~\eqref{eq:diag} is to be compared with the predictions of conventional
statistical mechanics ensembles, for a system in equilibrium with energy 
$E_I=\bra{\psi_I}\op{H}\ket{\psi_I}$ and total number of particles $N$. The canonical 
ensemble predicts
\begin{equation}
\expv{\hat{O}}_\text{CE} =\frac{1}{Z_\text{CE}}\sum_\alpha e^{-E_\alpha/k_BT} O_{\alpha\alpha},
\label{eq:CE}
\end{equation}
where $Z_\text{CE}=\sum_\alpha e^{-E_\alpha/k_BT}$, $T$ needs to be taken such that 
$E_I=Z^{-1}_\text{CE}\sum_\alpha e^{-E_\alpha/k_BT} E_\alpha$, and the sums run over all 
eigenstates of the Hamiltonian (with energy $E_\alpha$) in the sector with $N$ particles. 
The grand-canonical ensemble, on the other hand, predicts
\begin{equation}
\expv{\hat{O}}_\text{GE} =\frac{1}{Z_\text{GE}}\sum_\alpha e^{-(E_\alpha-\mu N_\alpha)/k_BT} O_{\alpha\alpha},
\label{eq:GE}
\end{equation}
where $Z_\text{GE}=\sum_\alpha e^{-(E_\alpha-\mu N_\alpha)/k_B T}$, $T$ and $\mu$ need to be taken such that 
$E_I=Z^{-1}_\text{GE}\sum_\alpha e^{-(E_\alpha-\mu N_\alpha)/k_BT} E_\alpha$ and 
$N=Z^{-1}_\text{GE}\sum_\alpha e^{-(E_\alpha-\mu N_\alpha)/k_BT} N_\alpha$, and the sums run over all 
eigenstates of the Hamiltonian (with energy $E_\alpha$ and number of particles $N_\alpha$).
The predictions of Eq.~\eqref{eq:CE} and Eq.~\eqref{eq:GE} in general agree in the thermodynamic limit.

Thermalization is then said to occur if, for sufficiently large systems, 
$\expv{\hat{O}}_\text{DE}\simeq\expv{\hat{O}}_\text{CE}\simeq\expv{\hat{O}}_\text{GE}$. Hence, the fact that 
thermalization occurs in isolated systems is surprising as $\expv{\hat{O}}_\text{DE}$ depends on 
the initial conditions through the projection of the initial state onto all the eigenstates of the Hamiltonian, 
while conventional statistical ensembles depend only on the initial conditions through $E_I$ and $N$. Since the 
energy distribution of the initial state $\ket{\psi_I}$ in the eigenstates of the final Hamiltonian is narrow 
(because of locality, see Ref.~\cite{rigol_dunjko_08} and its supplementary materials) 
and centered around $E_I$, similarly to the canonical and grand-canonical ensembles, then thermalization 
can be understood to occur because of ETH \cite{deutsch_91,srednicki_94,rigol_dunjko_08}. ETH states that in 
generic many-body systems $O_{\alpha\alpha}$ almost do not fluctuate between eigenstates that have similar 
energies, i.e., the eigenstates themselves already exhibit thermal behavior. 

Furthermore, it has been also proposed that one can define the entropy of the isolated 
system after the quench to be the diagonal entropy \cite{polkovnikov_11}
\begin{equation}
 S_d = -\sum_{\alpha} |C_\alpha|^2  \ln (|C_\alpha|^2),
\label{eq:diage}
\end{equation}
which satisfies all the thermodynamic properties required from an entropy. Indeed, this entropy has been recently
shown to be consistent with the microcanonical entropy for nonintegrable systems \cite{santos_polkovnikov_11},
and hence, for sufficiently large systems it is expected to agree with the entropy of the canonical ensemble
\begin{equation}
 S_\text{CE}=\ln Z_\text{CE} + \frac{E_I}{k_BT},
\label{eq:CEe}
\end{equation}
and with that of the grand-canonical ensemble
\begin{equation}
 S_\text{GE}=\ln Z_\text{GE} + \frac{E_I-\mu N}{k_BT},
\label{eq:GEe}
\end{equation}
up to subextensive corrections.

In general, in integrable systems such as the ones of interest in this work, the presence of a complete set 
of conserved quantities prevents thermalization \cite{rigol_dunjko_07,rigol_dunjko_08}. (The eigenstate 
thermalization hypothesis has been shown to fail in those systems \cite{rigol_dunjko_08,cassidy_clark_11}.) 
However, after relaxation, few-body observables can be described by means of a generalization of the Gibbs 
ensemble \cite{rigol_dunjko_07}, with a density matrix
\begin{equation}
\hat{\rho}_\text{GGE} = Z_\text{GGE}^{-1} e^{-\sum_n\lambda_n \hat{I}_n},
\label{eq:GGE}
\end{equation}
where $Z_\text{GGE}= \text{Tr} \left[ e^{-\sum_n\lambda_n \hat{I}_n} \right]$, \{$\hat{I}_n$\} are the 
conserved quantities, and $n=1,\ldots,L$. In our systems, \{$\hat{I}_n$\} are the occupation operators 
of the single-particle eigenstates of the noninteracting fermionic Hamiltonian to which hard-core bosons 
can be mapped. \{$\lambda_n$\} are the Lagrange multipliers, which are selected such that 
$\bra{\psi_I}\hat{I}_{n}\ket{\psi_I}=\text{Tr}(\hat{I}_{n}\hat{\rho}_\text{GGE})$. For hard-core bosons, 
they can be computed using the expression \cite{rigol_dunjko_07}
\begin{equation}
\lambda_n=\ln \left[\frac{1-\bra{\psi_I}\hat{I}_{n}\ket{\psi_I}}{\bra{\psi_I}\hat{I}_{n}\ket{\psi_I}}\right]
\label{eq:lagrangem}
\end{equation}
and $Z_{\text{GGE}}$ is then
\begin{equation}
Z_\text{GGE}= \prod_{n}(1+e^{-\lambda_{n}}).
\end{equation}

The fact that the GGE is able to predict expectation values of few-body observables after relaxation 
can be understood in terms of a generalized ETH \cite{cassidy_clark_11}. The idea in this case is that
eigenstates of the Hamiltonian that have similar values of the conserved quantities have 
similar expectation values of few-body observables. The GGE is then the ensemble that, within the 
full spectrum, selects a narrow set of states with the same distribution of conserved quantities that is 
fixed by the initial state. 

The GGE entropy is given by
\begin{equation}
 S_\text{GGE}=\ln Z_\text{GGE} + \sum_n\lambda_n \bra{\psi_I}\hat{I}_{n}\ket{\psi_I}.
\label{eq:GGEe}
\end{equation}
Furthermore, one can also define a canonical version of this generalized ensemble, with a density matrix
\begin{equation}
\hat{\rho}_\text{GCE} = Z_\text{GCE}^{-1} e^{-\sum_n\lambda_n \hat{I}_n},
\label{eq:GCE}
\end{equation}
for which only states with $N$ particles are considered when calculating traces. We keep $\lambda_n$ in the 
sector with $N$ particles to have the same values as within the GGE and take the partition function to be 
the trace over states with $N$ particles, $Z_\text{GCE}= \text{Tr} \left[ e^{-\sum_n\lambda_n \hat{I}_n} \right]_N$. 
The entropy of this ensemble can be computed as
\begin{equation}
S_\text{GCE}=\mbox{Tr}[\hat{\rho}_\text{GCE}\ln(\hat{\rho}_\text{GCE})]_N
\label{eq:GCEe}
\end{equation}
where, once again, only eigenstates of the Hamiltonian with $N$ particles contribute 
to the trace.

An interesting recent finding in Ref.~\cite{santos_polkovnikov_11} was that despite the fact that the
generalized ensembles do describe few-body observables in integrable systems after relaxation, their 
entropy is always greater than that of the diagonal ensemble, and the difference increases linearly with 
increasing system size. This means that an exponentially larger number of states contribute to the generalized
ensembles when compared to the diagonal one. The generalized ETH ensures that, despite having a much greater
number of states, the generalized ensembles predict the outcome of the realization dynamics. This is because the 
overwhelming majority of the states that contribute to the generalized ensembles have identical expectation 
values of few-body observables as the ones that contribute to the diagonal ensemble \cite{cassidy_clark_11}.
All these results are expected to be generic in integrable systems.

\begin{figure}[!t]
\begin{center}
 \includegraphics[width=0.485\textwidth]{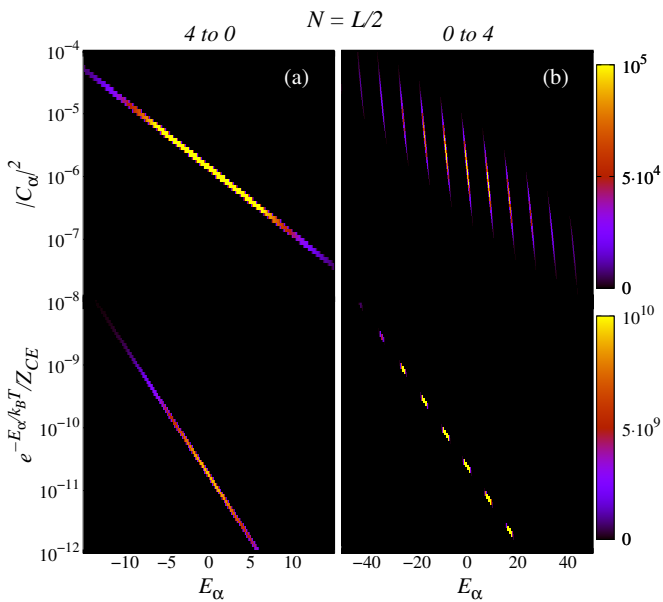}
\end{center}
\vspace{-0.6cm}
\caption{\label{fig:WeightsHalf} (Color online) Weights of the eigenstates of the final Hamiltonian in the 
diagonal (top half in both panels) and canonical (bottom half in both panels) ensembles, 
$|C_\alpha|^2$ and $e^{-E_\alpha/k_BT}/Z_\textrm{CE}$, respectively, for $L=36$ and $N=18$ 
(half filling). The panel on the left (a) depicts results for a quench from $A_I=4$ to $A_F=0$, 
and the panel on the right (b) for a quench from $A_I=0$ to $A_F=4$. In both cases, we select 
the initial state to be the ground state of Eq.~\eqref{eq:hamilt} for the given value of $A=A_I$. 
The color scale indicates the number of states, per unit area in the plot, that have a given weight.}
\end{figure}

In this work, instead, we focus on special classes of initial states that lead to expectation values of 
few-body observables that approach those in thermal equilibrium, despite integrability
\cite{rigol_muramatsu_06,cassidy_clark_11}. Since we know that 
ETH is not satisfied in those systems \cite{rigol_dunjko_08,cassidy_clark_11}, the fact that observables 
after relaxation approach thermal values then must be related to special properties of the overlaps $C_\alpha$ 
of the initial states with the eigenstates of the final Hamiltonians. Hence, we study the behavior of the 
$C_\alpha$'s in such systems. Hard-core bosons can be mapped onto noninteracting fermions, so one can generate 
the exponentially large Hilbert space of finite systems [whose size is $\binom{L}{N}$] without the need of 
diagonalizing the full Hamiltonians. Those many-body states are created as products of noninteracting fermionic 
eigenstates. They can be written as Slater determinants, in terms of fermionic creation operators 
$\hat{f}^{\dag}_{k}$, as 
\begin{equation}
\ket{\Psi_\alpha}=\prod^{N}_{l=1}\,\sum^L_{k=1}\,P_{k l}^{\alpha}\,\hat{f}^{\dag}_{k}\,|0 \rangle,
\end{equation}
and the same can be done for the initial state 
$\ket{\Psi_I}=\prod^{N}_{l=1}\sum^L_{k=1}P_{k l}^{0}\,\hat{f}^{\dag}_{k}\,|0 \rangle$. The overlap between the
initial state and the eigenstates of the final Hamiltonian can then be calculated numerically as the 
determinant of the product of two matrices \cite{rigol_muramatsu_05HCBb,he_rigol_11}
\begin{equation}
C_\alpha=\langle \Psi_\alpha \ket{\psi_I}=
\det\left[\left( \mathbf{P}^{\alpha} \right)^{\dag}\mathbf{P}^{0}\right],
\end{equation}
which, together will all the expressions presented previously, allow us to compute the energy distributions
and entropies in the diagonal, canonical, grand-canonical, and generalized ensembles.

\section{Overlaps} \label{sec:overlaps}

We first focus on the behavior of the weights $|C_\alpha|^2$ determined by the initial state and compare 
it with the one given by the canonical ensemble $e^{-E_{\alpha}/k_BT}/Z_\textrm{CE}$. Most of the results
reported in this manuscript are obtained from calculations for superlattices with period two. 
What that means is that in Eq.~\eqref{eq:hamilt}, 
\[
 V^{\textrm{ext}}_{j}=A (-1)^j.
\]
We will mainly focus on fillings (i) $N=L/2$ (half filling), for which observables after relaxation 
were seen to quickly approach the thermal predictions when the value of $A_I$ was increased and $A_F=0$, but no 
such thing was observed when $A_I=0$ and $A_F$ was increased \cite{rigol_muramatsu_06}, and (ii) $N=L/4$ 
(quarter filling), which does not exhibit an approach to the thermal predictions, like the one seen at half filling,
no matter the selected values of $A_I$ and $A_F$. Some results for trapped systems, related to the findings in 
Ref.~\cite{cassidy_clark_11}, will be reported in the following section. 

\begin{figure}[!t]
\begin{center}
 \includegraphics[width=0.485\textwidth]{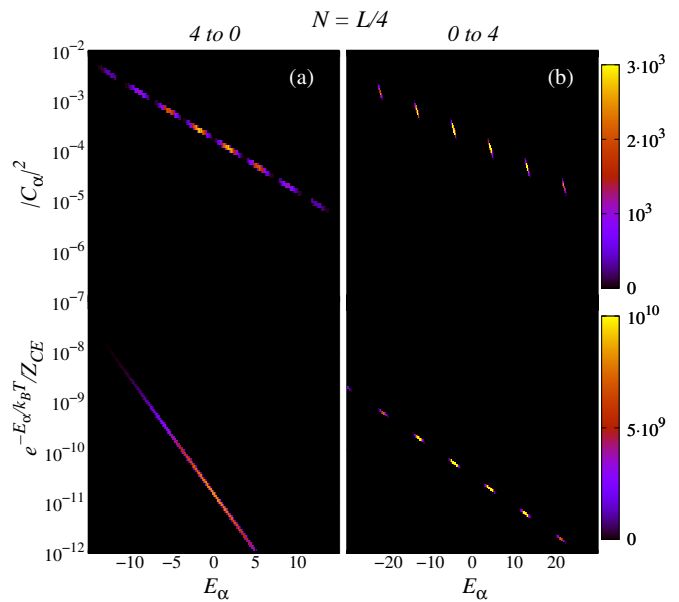}
\end{center}
\vspace{-0.6cm}
\caption{\label{fig:WeightsQuarter} (Color online) Same as Fig.~\ref{fig:WeightsHalf} but for $L=44$ and $N=11$ 
(quarter filling).}
\end{figure}

By comparing Eqs.~\eqref{eq:diag} and \eqref{eq:CE}, one may naively think that for those states
for which an approach to thermal expectation values was observed, the weights of the initial state 
in the eigenstates of the final Hamiltonian $|C_\alpha|^2$ may approach those of the canonical ensemble 
$e^{-E_\alpha/k_BT}/Z_\textrm{CE}$. In Fig.~\ref{fig:WeightsHalf}, we plot the values of $|C_\alpha|^2$ 
(top half in both panels) and $e^{-E_\alpha/k_BT}/Z_\textrm{CE}$ (bottom half in both panels) for quenches 
from the ground state of a superlattice ($A_I=4$) to the homogeneous lattice ($A_F=0$) (a)
and from the ground state of the homogeneous lattice ($A_I=0$) to the superlattice ($A_F=4$) (b).

Figure \ref{fig:WeightsHalf} clearly shows that the actual values of $|C_\alpha|^2$ differ not only 
quantitatively (several orders of magnitude) from those of $e^{-E_\alpha/k_BT}/Z_\textrm{CE}$ 
but also qualitatively different for both quenches, as the former exhibit a slower decay with the energy 
of the eigenstates. No convergence between the values of $|C_\alpha|^2$ and $e^{-E_\alpha/k_BT}/Z_\textrm{CE}$ 
is observed as $A_I$ and $A_F$ are changed (not shown). In Fig.~\ref{fig:WeightsHalf}, we also provide 
information about the number of states, per unit area in the plot, that have a given weight within 
in each ensemble (color scale). For the quench from $A_I=4$ to $A_F=0$, one can see in Fig.~\ref{fig:WeightsHalf}(a) 
(top half) that the number of states with nonzero values of $|C_\alpha|^2$ continuously 
increases as the energy increases and reaches a maximum around the center of the spectrum, where the 
density of states is largest. A similar behavior can be seen within the canonical ensemble in 
Fig.~\ref{fig:WeightsHalf}(a) (bottom half). For the quenches from $A_I=0$ to $A_F=4$, on the 
other hand, there are isolated islands with nonzero weights both in the diagonal and canonical ensembles 
[Fig.~\ref{fig:WeightsHalf}(b)]. This is because the many-body spectrum exhibits bands 
of eigenstates separated by gaps, which are determined by the value of $A_F$. Such a behavior can be 
straightforwardly understood from the single-particle band structure. In the periodic case, a reasonably 
good approximation for large systems with open boundary conditions, the latter exhibits two bands given 
by the expression
\begin{equation}
\epsilon_{\pm}(k) = \pm \sqrt{ 4 t^2 \cos^2(ka) + A^2} ,
\end{equation}
where ``$+$'' denotes the upper band and ``$-$'' the lower band and $k$ is the single particle 
momentum. Depending on which values of $k$ are occupied in the many-body state, the bands seen in 
Fig.~\ref{fig:WeightsHalf}(b) form.

Results for the same quenches as in Fig.~\ref{fig:WeightsHalf}, but for quarter-filled systems,
are presented in Fig.~\ref{fig:WeightsQuarter}. The latter are qualitatively similar to the former
in everything, except for the behavior of the number of eigenstates with nonzero values of 
$|C_\alpha|^2$ in the quenches from $A_I=4$ to $A_F=0$ [top half in Fig.~\ref{fig:WeightsQuarter}(a)].
At quarter filling, when $A_I=4$, the initial state imprints a modulation on the number of 
eigenstates with nonzero $|C_\alpha|^2$ (note that the spectrum in the final Hamiltonian, 
when $A_F=0$, has no gaps). That modulation is not present for the quenches at half filling 
[top half in Fig.~\ref{fig:WeightsHalf}(a)] and, as expected (because of the continuous spectrum 
of the final Hamiltonian), it is not present in the canonical results in the bottom half of 
Fig.~\ref{fig:WeightsQuarter}(a).

The fact that the weights in the diagonal and canonical (or any other) ensemble differ
from each other is generic for integrable and nonintegrable systems \cite{rigol_10} and, as 
such, need not preclude thermalization. After all, the weights with which eigenstates of the 
Hamiltonian contribute to the canonical and microcanonical ensembles also differ. 
The relevant quantity to compare different ensembles is the energy density $\rho(E)$, which 
is equal to the sum of the weights studied in Figs.~\ref{fig:WeightsHalf} and \ref{fig:WeightsQuarter}, 
over a given energy window $\delta E$, divided by $\delta E$. By construction, the integral 
of this quantity over the full energy spectrum is normalized to 1. ($\delta E$ 
needs to be selected in such a way that the results for the energy density are independent 
of its actual value.) The energy density depends not only on the weights but also on the density 
of states, and tells us which part of the spectrum is the one that contributes the most to 
the ensemble averages.

\begin{figure}[!t]
\begin{center}
 \includegraphics[width=0.48\textwidth]{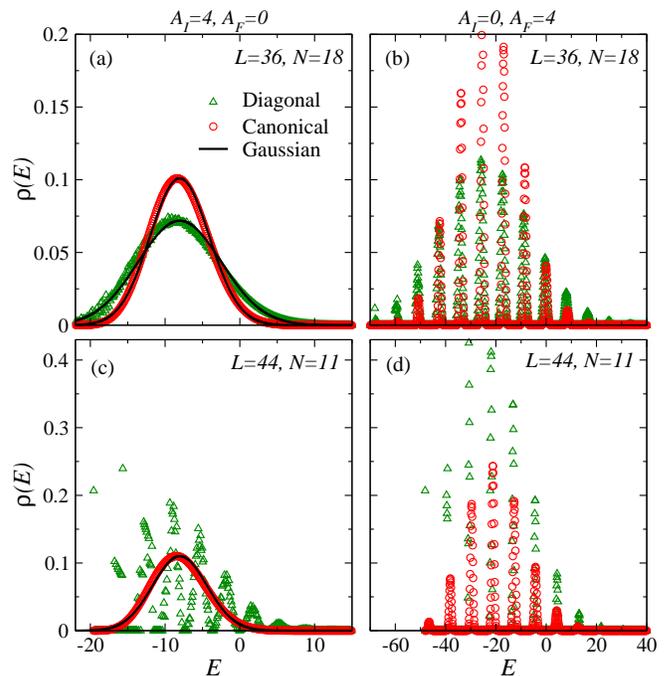}
\end{center}
\vspace{-0.6cm}
\caption{\label{fig:EnergyDensity} (Color online) Energy density $\rho(E)$ for the quenches depicted in 
Figs.~\ref{fig:WeightsHalf} and \ref{fig:WeightsQuarter}. Results are presented for the case 
$A_I=4$, $A_F=0$ in the left panels [(a) and (c)] and for $A_I=0$, $A_F=4$ in the right panels 
[(b) and (d)] and for systems at half filling [(a) and (b)] and at quarter filling [(c) and (d)]. 
$\rho(E)$ is reported for the diagonal and canonical ensembles, and, when appropriate, we have fitted
the results to a Gaussian (continuous lines in the plots). In all cases $\delta E=0.1$.}
\end{figure}

In Fig.~\ref{fig:EnergyDensity}, we present $\rho(E)$ for the quenches for which the weights
of the diagonal and canonical ensembles were reported in Figs.~\ref{fig:WeightsHalf} and 
\ref{fig:WeightsQuarter}. As expected, the energy density in the canonical ensemble is very close
to a Gaussian $\rho(E)=(\sqrt{2 \pi}\delta E)^{-1} e^{-(E-E_I)^2/(2 \delta E^2)}$ in all cases.
For the quenches from $A_I=0$ to $A_F=4$, the Gaussian is cut by the bands described previously. 

In diagonal ensemble, $\rho(E)$ has been recently 
shown to be very well described by a Gaussian for nonintegrable systems and sparse (very different from 
Gaussian) in integrable systems \cite{santos_polkovnikov_11}. We find the latter to be generic 
for our quenches, as shown in Figs.~\ref{fig:EnergyDensity}(c), \ref{fig:EnergyDensity}(d), and, 
maybe less evident but still true, Fig.~\ref{fig:EnergyDensity}(b). Surprisingly, we find that for 
quenches from $A_I\neq0$ to $A_F=0$ at half filling, the energy density in the diagonal ensemble 
approaches a Gaussian as the value of $A_I$ is increased. See Fig.~\ref{fig:EnergyDensity}(a) 
for $A_I=4$ and $A_F=0$. This highlights the special character of this class of initial states 
and will be analyzed more quantitatively in the following sections.

\begin{figure}[!t]
\begin{center}
 \includegraphics[width=0.48\textwidth]{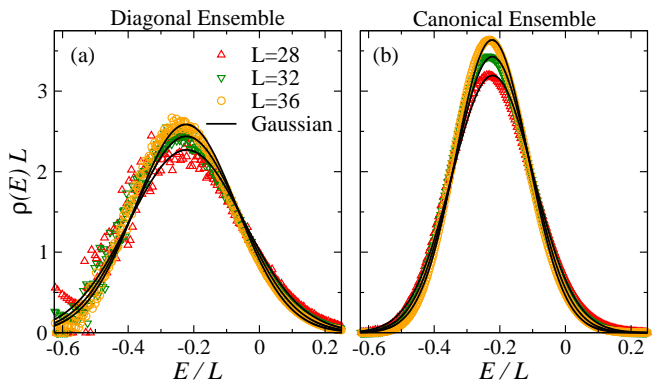}
\end{center}
\vspace{-0.6cm}
\caption{\label{fig:EnergyDensity_Scaling} (Color online) Scaling of the energy density in the diagonal (a)
and canonical ensembles (b) with increasing system size. Results are reported for the quenches 
from $A_I=4$ to $A_F=0$ at half filling. Continuous lines depict the result of the fit of each 
data set to a Gaussian.}
\end{figure}

A remark is in order on the scaling of the plots shown in Fig.~\ref{fig:EnergyDensity} with increasing 
system size. For the canonical ensemble, it is known that the width of $\rho(E)$, relative to the full
width of the spectrum, vanishes in the thermodynamic limit. The question is then what happens for the diagonal 
ensemble. On general grounds, for Hamiltonians containing only finite-range terms, it was shown in 
Ref.~\cite{rigol_dunjko_08} (supplementary materials) that the width of $\rho(E)$,  
relative to the full width of the spectrum, also vanishes in the thermodynamic limit. 
The scaling of the width of $\rho(E)$ depends in this case on the nature of the quench
\cite{rigol_dunjko_08}. In Fig.~\ref{fig:EnergyDensity_Scaling}, we show a finite size scaling for 
$\rho(E)$ in the diagonal (a) and canonical (b) ensembles in the quenches from $A_I=4$ to $A_F=0$. 
These results are consistent with the vanishing of the width of $\rho(E)$, relative to the width of 
the spectrum, as the system size is increased.

\section{Entropies} \label{sec:entropies}

In the previous section, we have shown that the energy distribution in quenches whose initial states are 
the ground state of half-filled systems with a superlattice ($A_I\neq0$) can be well described by a Gaussian, 
typical of thermal states, as the value of $A_I$ is increased. However, at least for the finite systems we 
can solve numerically, we showed that such a Gaussian like energy distribution clearly differs from that of the 
canonical ensemble. In this section, we use the entropies, including the diagonal entropy $S_d$ 
\cite{polkovnikov_11,santos_polkovnikov_11}, as a way to quantify the scaling of the energy distributions 
in all ensembles as the system size is increased. In Ref.~\cite{santos_polkovnikov_11}, it was already 
shown that the diagonal entropy in integrable systems increases nearly linearly with system size, 
demonstrating its additive character.

\begin{figure}[!h]
\begin{center}
 \includegraphics[width=0.485\textwidth]{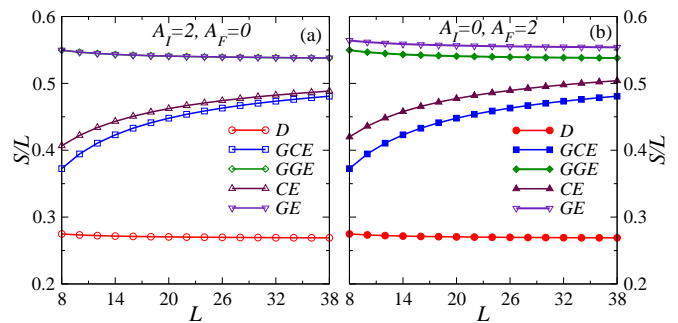}
\end{center}
\vspace{-0.7cm}
\caption{\label{fig:EntropyQuenchA02} (Color online) Entropy per site vs. $L$ in quenches from the ground state
of a superlattice with $A_I=2$ to the homogeneous lattice $A_F=0$ (a) and from the homogeneous lattice
$A_I=0$ to the superlattice with $A_F=2$ (b). For both quenches, we show results for the diagonal $S_d$,
canonical $S_\textrm{CE}$, grand-canonical $S_\textrm{GE}$, generalized canonical $S_\textrm{GCE}$, and 
generalized grand-canonical $S_\textrm{GGE}$ entropies. The systems are at half filling $N=L/2$.}
\end{figure}

In Fig.~\ref{fig:EntropyQuenchA02}, we show the entropy per site for two different quenches in 
half-filled systems, with increasing system size. For both quenches, one can see that all entropy per site 
plots tend to saturate to a constant value with increasing $L$, making evident the additivity
of this observable in all ensembles. Another result that is apparent from those plots is that $S_d$ is 
smaller than all other entropies, and it seems that it will remain that way in the thermodynamic limit,
as noted in Ref.~\cite{santos_polkovnikov_11}. An important difference between the behavior of the 
entropies for a quench from the superlattice to the homogeneous lattice [Fig.~\ref{fig:EntropyQuenchA02}(a)]
and the quench from the homogeneous lattice to the superlattice [Fig.~\ref{fig:EntropyQuenchA02}(b)] is that, 
in the former, the entropy of the GGE and the grand-canonical ensemble are nearly identical and the entropies 
of the GCE and the canonical ensemble approach each other with increasing system size. In the latter quench,
the entropies of the GGE and the grand-canonical ensemble differ from each other and their difference
is seen to remain constant as the system size is increased. $S_\textrm{GCE}$ and $S_\textrm{CE}$ approach 
each other as the system size increases, but their difference is clearly larger than for the $A_I=2$ to $A_F=0$
quench.

\begin{figure}[!t]
\begin{center}
 \includegraphics[width=0.485\textwidth]{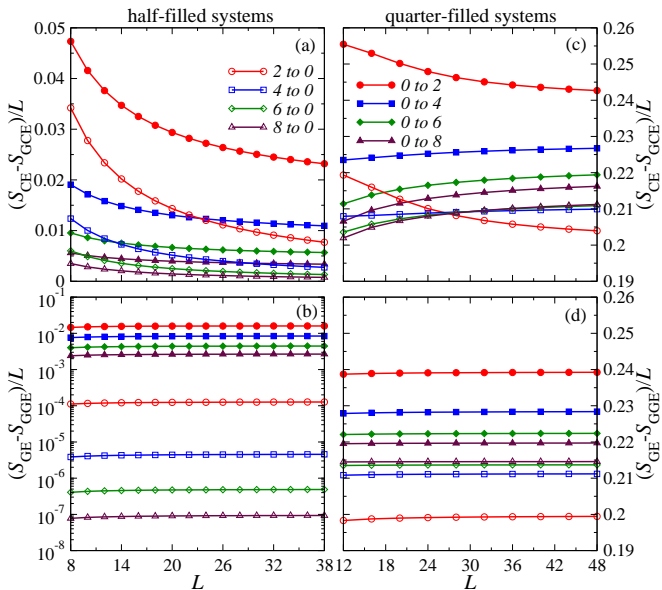}
\end{center}
\vspace{-0.7cm}
\caption{\label{fig:EntropyScaling} (Color online) Difference between the entropy of the canonical 
ensemble and the GCE [(a) and (c)] and between the grand-canonical ensemble and the GGE [(b) and (d)] 
for quenches at half filling [(a) and (b)] and quarter filling [(c) and (d)]. In all panels, the results 
for the quenches from the superlattice to the homogeneous lattice are depicted using open symbols, 
while the ones from the homogeneous lattice to the superlattice are depicted using solid symbols. 
For the former quenches, results are reported for $A_I=2$ and $A_F=0$, $A_I=4$ and $A_F=0$, 
$A_I=6$ and $A_F=0$, and $A_I=8$ and $A_F=0$, and, for the latter, results are reported for 
$A_I=0$ and $A_F=2$, $A_I=0$ and $A_F=4$, $A_I=0$ and $A_F=6$, and $A_I=0$ and $A_F=8$. In the legend,
we use the notation ``$A_I$ to $A_F$'' to label the plots.}
\end{figure}

In order to quantify the observations above for different quenches and fillings, 
in Fig.~\ref{fig:EntropyScaling} we plot the scaling of $(S_\textrm{CE}-S_\textrm{GCE})/L$ and
$(S_\textrm{GE}-S_\textrm{GGE})/L$ with system size. The left panels
[Figs.~\ref{fig:EntropyScaling}(a) and \ref{fig:EntropyScaling}(b)] depict the results at 
half filling. Figure \ref{fig:EntropyScaling}(a) shows that for any given pair
$A_I=x\rightarrow A_F=0$ and $A_I=0\rightarrow A_F=x$, where $x=2,4,6,8$, the difference
$(S_\textrm{CE}-S_\textrm{GCE})/L$ saturates at greater values for the quenches starting 
from the homogeneous lattice than for those starting from the superlattice (which, 
for the lattice sizes shown, still keep decreasing as the system size is increased). The 
difference $(S_\textrm{GE}-S_\textrm{GGE})/L$, in Fig.~\ref{fig:EntropyScaling}(b), 
exhibits and even more remarkable behavior. It does not change with increasing system 
size, and it can be seen to be orders of magnitude smaller for the quenches starting from 
the superlattice when compared to those starting from the homogeneous system. The 
difference $(S_\textrm{GE}-S_\textrm{GGE})/L$ quickly approaches zero as the value 
$A_I$ in the superlattice is increased. This is exactly the same behavior that was 
observed in Ref.~\cite{rigol_muramatsu_06} for the difference between the expectation 
value of the momentum distribution function ($n_k$) in the grand-canonical ensemble 
and that of the time average in the time evolving state. (The latter can be reproduced 
using the GGE.) 

Hence, we can conclude that for the particular class of initial states 
in Ref.~\cite{rigol_muramatsu_06}, where quenches starting from the ground state of a system 
with a superlattice lead to expectation values of $n_k$ that approach those in thermal 
equilibrium as $A_I$ was increased, the sets of states that contribute to the grand-canonical
ensemble and the GGE become increasingly similar to each other. Since the GGE describes 
observables in the integrable system after relaxation \cite{rigol_dunjko_07,cassidy_clark_11}, 
thermal ensembles then will also provide a very good estimate for those observables as $A_I$ is
increased. From the results in Fig.~\ref{fig:EntropyScaling}(b), it is important to stress that
the entropies in the grand-canonical ensemble and the GGE do not approach each other, for a fixed 
value of $A_I$, as the system size is increased. 

In Figs.~\ref{fig:EntropyScaling}(c) and \ref{fig:EntropyScaling}(d), we show results for an 
identical set of quenches as the one in Figs.~\ref{fig:EntropyScaling}(a) and \ref{fig:EntropyScaling}(b) 
but for systems at quarter filling. For all quenches at quarter filling, one can see that the 
differences between the entropy in the standard ensembles and in the generalized ones is orders 
of magnitude larger than for the quenches at half filling. The differences between the two are maximal 
for the quenches with $A_I\neq0$. The behavior with changing system size is, however, similar to the one in 
the systems at half filling. Hence, by comparing all panels in Fig.~\ref{fig:EntropyScaling}, one can 
further see that there is something special about the quenches starting from the half-filled 
superlattice.

\begin{figure}[!t]
\begin{center}
 \includegraphics[width=0.4\textwidth]{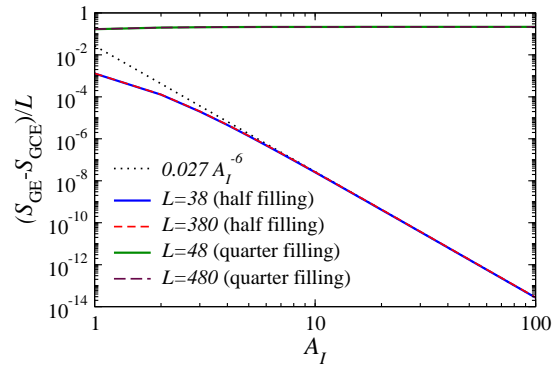}
\end{center}
\vspace{-0.7cm}
\caption{\label{fig:EntropiesvsAI} (Color online) Difference between the entropy of the grand-canonical 
ensemble and the GGE for quenches at half filling (two lower curves) and quarter filling 
(two upper curves) vs. $A_I$. In all cases $A_F=0$. The dotted line depicts a power-law fit 
to the large $A_I$ results at half filling. For each quench, results for two different system 
sizes are presented.}
\end{figure}

As discussed in Refs.~\cite{rigol_muramatsu_06,rousseau_arovas_06}, the ground state 
of half-filled systems in a superlattice is insulating and, as the value of $A$ increases, 
its wavefunction approaches that of a trivial Fock state [a product state of empty 
(low chemical potential) and occupied (high chemical potential) sites]. 
In Ref.~\cite{rigol_muramatsu_06}, it was shown that the one-particle correlation length $\xi$ 
decays as a power law $\xi/a\sim 1/\sqrt{A/t}$ for large values of $A/t$ ($A/t\gtrsim 4$). 
In Fig.~\ref{fig:EntropiesvsAI}, we show how $(S_\textrm{GE}-S_\textrm{GGE})/L$ decreases as
$A_I$ increases. Here again, we find a power-law decay for large values of $A_I$, where 
$(S_\textrm{GE}-S_\textrm{GGE})/L\sim 1/A_I^6$. This large exponent explains the fast 
reduction of $(S_\textrm{GE}-S_\textrm{GGE})/L$ seen in Fig.~\ref{fig:EntropyScaling} 
when $A_I$ was increased. 

Since for large values of $A_I$ the initial states are nearly uncorrelated states ($\xi\rightarrow0$), their 
overlaps with the eigenstates of the final Hamiltonian can be understood to be random and 
constrained only by energy conservation. This helps in understanding the origin of Gaussian energy 
distribution observed in the previous section and the closeness of the generalized ensemble entropies 
to those of the standard ensembles as $A_I$ is increased. In Fig.~\ref{fig:EntropyScaling}, we 
also present results for the quenches at quarter filling, where $(S_\textrm{GE}-S_\textrm{GGE})/L$ 
is seen to saturate to a finite value when $A_I$ is increased. For both fillings, and the 
system sizes depicted in that figure, finite-size effects can be seen to be negligible.

Confirmation of the conclusions above can be obtained if one realizes that a similar 
argument applies to the systems discussed in Ref.~\cite{cassidy_clark_11}. There, the 
initial state was selected to be the ground state of a trapped system, where [Eq.~\eqref{eq:hamilt}]
\[
 V^{\textrm{ext}}_{j}= V (j-L/2)^2
\]
is a harmonic trapping potential and the evolution was followed after the trap potential $V$ 
was turned off, {\it i.e.}, $V_I\neq 0$ and $V_F=0$. For a fixed number of particles, as $V_I$  
increases, a Mott insulator (Fock state for hard-core bosons) with density $n=1$ forms in the center 
of the trap. When initial states containing such Mott insulating domains were used for the time 
evolution, the difference between the momentum distribution function in the diagonal ensemble and standard 
ensembles of statistical mechanics was seen to decrease \cite{cassidy_clark_11}. 

\begin{figure}[!t]
\begin{center}
 \includegraphics[width=0.485\textwidth]{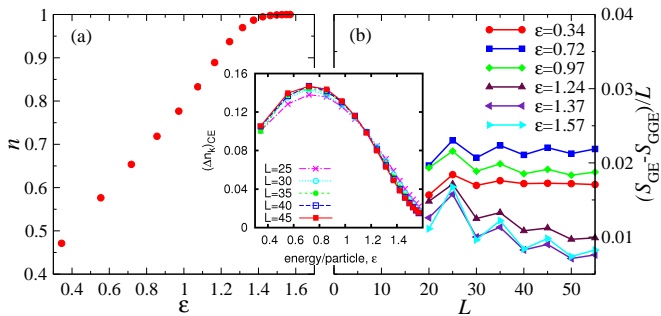}
\end{center}
\vspace{-0.7cm}
\caption{\label{fig:Trap} (Color online) (a) Density in the center of the trap as a function of the 
excitation energy per particle $\varepsilon$, which is changed by increasing $V_I$ in a system 
with 50 lattice sites and 10 particles. (b) Difference between the entropy of the grand-canonical 
ensemble and the GGE vs. $L$ for systems with different excitation energy per particle.
(Inset): Integrated relative difference between $n_k$ in the diagonal and canonical ensembles 
(see text) vs. the excitation energy per particle. Results are presented for different system 
sizes, $L=25$, 30, \ldots, 45 \cite{cassidy_clark_11}.}
\end{figure}

In the inset in Fig.~\ref{fig:Trap}, we show the results obtained in Ref.~\cite{cassidy_clark_11} 
for the integrated difference between the predictions of the diagonal and canonical ensembles for $n_k$ 
\[
(\Delta n_k)_\text{CE}=\frac{\sum_k|\expv{\hat{n}_k}_\text{DE} - 
\expv{\hat{n}_k}_\text{CE}|}{\sum_k\expv{\hat{n}_k}_\text{DE}},
\]
as a function of the excitation energy per particle 
\[
 \varepsilon=\frac{E_I-E_G}{N},
\]
where $E_G$ is the ground-state energy of the final (homogeneous) Hamiltonian. The excitation energy 
increases by increasing $V_I$, while keeping $L$ and $N$ constant \cite{cassidy_clark_11}. 

The density in the center of the trap (in the initial state) versus the excitation energy is plotted in 
Fig.~\ref{fig:Trap}(a). There, one can see that $(\Delta n_k)_\text{CE}$ (in the inset) is smallest 
and keeps decreasing when the density in the center of the trap approaches or becomes 
equal to 1, i.e., when an increasingly large portion of the system comes close or becomes a Fock state.

The scaling of the difference between the entropies in grand-canonical ensemble and the GGE is shown in 
Fig.~\ref{fig:Trap}(b) for different excitation energies. Similarly to the results for the superlattice 
systems, that difference is seen to be smallest (and decreasing with increasing system size in this case) 
for the systems whose initial states are closest to Fock states. Hence, once again, a special class of 
initial states is seen to produce increasingly ``thermal-like'' observables and generalized ensembles.

\section{Conserved quantities} \label{sec:conserved}

Conserved quantities play a fundamental role in the dynamics and description after relaxation
of integrable systems. The latter follows from the evidence that generalized ensembles are able 
to describe observables after equilibration while standard statistical ensembles are, in general, not
\cite{rigol_dunjko_07,rigol_muramatsu_06,cazalilla_06,calabrese_cardy_07a,cramer_dawson_08,
barthel_schollwock_08,eckstein_kollar_08,kollar_eckstein_08,iucci_cazalilla_09,fioretto_mussardo_10,
iucci_cazalilla_10,cassidy_clark_11, calabrese_essler_11,cazalilla_iucci_11}. Hence, a distinctive 
behavior is expected of the distribution of the conserved quantities for those initial states for 
which observables after relaxation approach thermal values. In this section, 
we study the behavior of the conserved quantities in those and other cases analyzed in the previous 
sections.

As explained in Sec.~\ref{sec:models}, the expectation values of the conserved quantities in hard-core 
boson systems can be straightforwardly computed because they are the occupation of the eigenstates of 
the noninteracting fermionic Hamiltonian (there are $L$ of those) to which hard-core bosons can be 
mapped. As such, they can be calculated within the GGE (identical to those of the initial state and 
the diagonal ensemble by construction) and in the grand-canonical ensemble, for very large lattices. 
In the grand-canonical ensemble, the occupation of the conserved quantities is dictated 
by the Fermi distribution
\begin{equation}
 \expv{\hat{I}_n}_\text{GE} =\frac{1}{e^{(\epsilon_n-\mu)/k_BT}+1},
\label{eq:fermio}
\end{equation}
where $\epsilon_n$ are the single-particle eigenenergies.

\begin{figure}[!t]
\begin{center}
 \includegraphics[width=0.485\textwidth]{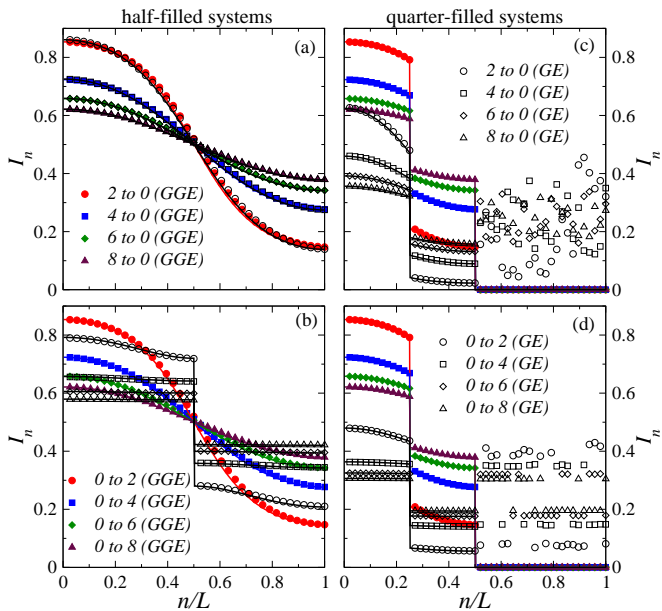}
\end{center}
\vspace{-0.7cm}
\caption{\label{fig:ConservedQuantities} (Color online) Expectation value of the conserved quantities in 
quenches from $A_I\neq0$ to $A_F=0$ [(a) and (c)] and from $A_I=0$ to $A_F\neq0$ [(b) and (d)] for 
systems at half filling [(a) and (b)] and systems at quarter filling [(c) and (d)]. The conserved quantities 
are ordered from the highest to the lowest occupied in the initial state. Solid symbols 
depict the results of the GGE (conserved quantities in the initial state)
and open symbols depict the results of the grand-canonical ensemble. The results 
denoted by symbols (lines) correspond to systems with 38 (380) sites in the half-filled case 
[(a) and (b)] and with 48 (480) sites in the quarter-filled case [(c) and (d)]. Note that 
for the smallest system sizes depicted here (the largest analyzed in the previous sections) finite-size 
effects for the conserved quantities already are negligible. They exhibit an almost perfect overlap
with the results in systems 10 times larger. Results are reported for quenches between 
$A_I=2$ and $A_F=0$, $A_I=4$ and $A_F=0$, $A_I=6$ and $A_F=0$, and $A_I=8$ and $A_F=0$ and between
$A_I=0$ and $A_F=2$, $A_I=0$ and $A_F=4$, $A_I=0$ and $A_F=6$, and $A_I=0$ and $A_F=8$. In the legend,
we use the notation ``$A_I$ to $A_F$'' to label the plots.}
\end{figure}

In Figs.~\ref{fig:ConservedQuantities}(a) and \ref{fig:ConservedQuantities}(b), we depict 
the conserved quantities in the GGE (initial state) and the grand-canonical ensemble for quenches 
at half filling from the ground state in a superlattice to the homogeneous lattice 
[Fig.~\ref{fig:ConservedQuantities}(a)] and from the ground state of the homogeneous lattice to 
the superlattice [Figs.~\ref{fig:ConservedQuantities}(b)]. (The conserved quantities are ordered 
from the highest to the lowest occupied in the initial state.) A clear contrast can be seen between 
those two panels. In Fig.~\ref{fig:ConservedQuantities}(a) the results for the GGE and grand-canonical 
ensemble are almost indistinguishable from each other while in Fig.~\ref{fig:ConservedQuantities}(b) 
they differ from each other markedly. This behavior does not change with increasing system 
size as, in the same figure, depicted as continuous lines, we also report results for lattices 10
times larger than those used for the calculations depicted as symbols. Qualitatively, these 
results are similar to those obtained in Ref.~\cite{cassidy_clark_11} for the trapped systems 
analyzed in the previous section. 

Insights into the contrast between the results in Fig.~\ref{fig:ConservedQuantities}(a) and 
Fig.~\ref{fig:ConservedQuantities}(b) can be gained if one notices that the distribution of 
conserved quantities for the quenches from and to the superlattice are smooth and identical 
when $A_I$ in the former is equal to $A_F$ in the latter. This immediately helps one
understand why the grand-canonical ensemble prediction of the conserved quantities can match 
that of the quenches from the superlattice ($A_I\neq0$) to the homogeneous lattice ($A_F=0$) 
but not that of the quenches from the homogeneous lattice ($A_I=0$) to the superlattice ($A_F\neq0$). 
In the former, the final system exhibits no gaps ($A_F=0$) and the conserved quantities
[the Fermi distribution, see Eq.~\eqref{eq:fermio}] can be a smooth function of $n$ at finite 
temperatures, while in the latter the system is gapped ($A_F\neq0$) 
and, hence, a discontinuity must occur in the Fermi distribution at the gap position [as seen in 
Fig.~\ref{fig:ConservedQuantities}(b) for $n/L=0.5$].

The results reported in Figs.~\ref{fig:ConservedQuantities}(c) and \ref{fig:ConservedQuantities}(d) for 
systems at quarter filling, are qualitatively similar to those in Fig.~\ref{fig:ConservedQuantities}(b). 
For all quenches, the conserved quantities in the initial state differ substantially from those predicted 
by the grand-canonical ensemble. The differences can be noted to be particularly large if one realizes that, 
for many conserved quantities, the initial state has a zero expectation value while the grand-canonical ensemble 
predicts nonzero, and large, values. This helps in understanding the large differences seen in the 
previous section between the entropies in the generalized and standard ensembles for the quenches at 
quarter filling. Once again, the behavior of the conserved quantities in thermal equilibrium is dictated by 
the Fermi distribution and can be understood given the gapless or gapped nature of the spectrum of the final system.

\begin{figure}[!t]
\begin{center}
 \includegraphics[width=0.485\textwidth]{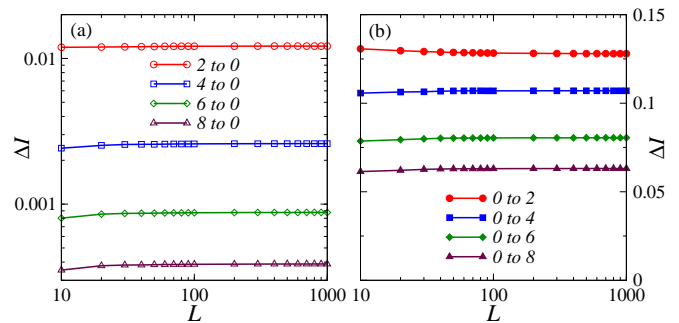}
\end{center}
\vspace{-0.7cm}
\caption{\label{fig:ScalingConservedQuantities} (Color online) Integrated differences between the conserved quantities
in the GGE (initial state) and in the grand-canonical ensemble (see text) vs. $L$ for half-filled systems
in quenches from $A_I\neq0$ to $A_F=0$ (a) and from $A_I=0$ to $A_F\neq0$ (b). Results are reported for 
quenches between $A_I=2$ and $A_F=0$, $A_I=4$ and $A_F=0$, $A_I=6$ and $A_F=0$, and $A_I=8$ and $A_F=0$ in (a); 
and, between $A_I=0$ and $A_F=2$, $A_I=0$ and $A_F=4$, $A_I=0$ and $A_F=6$, and $A_I=0$ and $A_F=8$, in (b). 
In the legend, we use the notation ``$A_I$ to $A_F$'' to label the plots.}
\end{figure}

Figure \ref{fig:ScalingConservedQuantities} depicts how the difference between the conserved quantities 
in the initial state and the grand-canonical ensemble, given by the integrated relative difference
\[
\Delta I=\frac{\sum_n|\expv{\hat{I}_n}_\text{GGE}-\expv{\hat{I}_n}_\text{GE}|}{\sum_n\expv{\hat{I}_n}_\text{GGE}},
\]
behaves as the system size increases. The results presented, 
for half-filled systems in quenches from a superlattice potential ($A_I\neq0$ and $A_F=0$) in
Fig.~\ref{fig:ScalingConservedQuantities}(a) and to a superlattice potential ($A_I=0$ and $A_F\neq0$) in
Fig.~\ref{fig:ScalingConservedQuantities}(b), show more quantitatively that the results in 
Fig.~\ref{fig:ConservedQuantities} do not change with increasing system size. 

\begin{figure}[!t]
\begin{center}
 \includegraphics[width=0.4\textwidth]{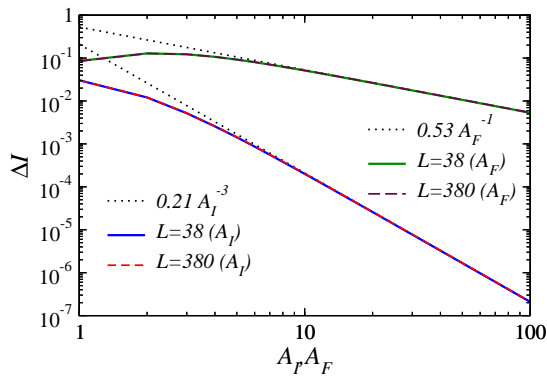}
\end{center}
\vspace{-0.7cm}
\caption{\label{fig:ConservedQuantitiesvsAI} (Color online) Integrated differences between the conserved 
quantities in the GGE and in the grand-canonical ensemble. Results are reported for quenches
where (i) $A_I\neq0$ and $A_F=0$ (two bottom curves) vs. $A_I$ and 
(ii) $A_I=0$ and $A_F\neq0$ (two upper curves) vs. $A_F$ and for two different system sizes.
The dotted lines depict a power-law fits to the large $A_I$, $A_F$ results. }
\end{figure}

Figure \ref{fig:ScalingConservedQuantities} also makes evident that there is a big quantitative 
difference between $\Delta I$ in the systems whose initial state is the ground state in the superlattice 
[Fig.~\ref{fig:ScalingConservedQuantities}(a)] and those whose initial state is the ground state 
of the homogeneous lattice [Fig.~\ref{fig:ScalingConservedQuantities}(b)]. This is better seen in 
Fig.~\ref{fig:ConservedQuantitiesvsAI}, where $\Delta I$ is plotted versus $A_I$ for the former case and 
versus $A_F$ for the latter. In both cases, we find power-law decays, which are $\sim 1/A_I^3$ when 
the initial state was created for $A_I\neq 0$ and $\sim 1/A_F$ when the final Hamiltonian has 
$A_F\neq0$. It is important to stress that while increasing $A_I$ does not qualitatively change
the time dynamics of observables of interest, increasing $A_F$ does \cite{rigol_muramatsu_06}. In 
the latter case the damping (relaxation) of the observables is inhibited \cite{rigol_muramatsu_06}, 
so the assumption that observables relax to time independent values breaks down. 

Finally, from Figs.~\ref{fig:ScalingConservedQuantities} and \ref{fig:ConservedQuantitiesvsAI}, we should 
emphasize once again that, complementary to the behavior seen for $(S_\textrm{GE}-S_\textrm{GGE})/L$ 
in the previous section, the scaling of $\Delta I$ versus $L$ is similar for both types of quenches, namely, 
any finite value of $A_I$ (if $A_F=0$) or $A_F$ (if $A_I=0$) leads to a finite $\Delta I$ in the thermodynamic 
limit. The difference between those quenches resides in the actual values of $\Delta I$ and 
their behavior with changing $A_I$ or $A_F$.

On the basis of those results we can now understand that, for the classes of initial states
in Refs.~\cite{rigol_muramatsu_06,cassidy_clark_11} for which observables after relaxation
approached thermal values, the control parameter used tuned the 
distribution of conserved quantities to approach thermal values (resulting in generalized 
ensembles that, for those states, approach thermal ensembles). This behavior, however, should not 
be confused with thermalization as understood for nonintegrable systems. For the latter, the 
difference between observables after relaxation and the predictions of statistical mechanics 
ensembles is expected to vanish in the thermodynamic limit, while, for the special classes of 
initial states that we have studied here for integrable systems, such a difference remains 
finite in the thermodynamic limit for any selected (finite) value of the control parameter.

To conclude, there is an important distinction to be made about the generalized ensembles when compared with 
standard ensembles of statistical mechanics. In the latter, the conserved quantities (energy, momentum, 
angular momentum, etc.) are additive and their number is $\sim 1$. In the generalized ensembles, the 
conserved quantities are, strictly speaking, not additive and their number, in the integrable 
systems considered here, is $\sim L$. The fact that they are not additive can be immediately seen 
in Fig.~\ref{fig:ConservedQuantities}, where, after making the system size 10 times larger, the value 
of the conserved quantities does not change. Instead, their number increased by a factor of 10
(that is the reason for plotting the conserved quantities as functions of $n/L$).

\begin{figure}[!h]
\begin{center}
 \includegraphics[width=0.485\textwidth]{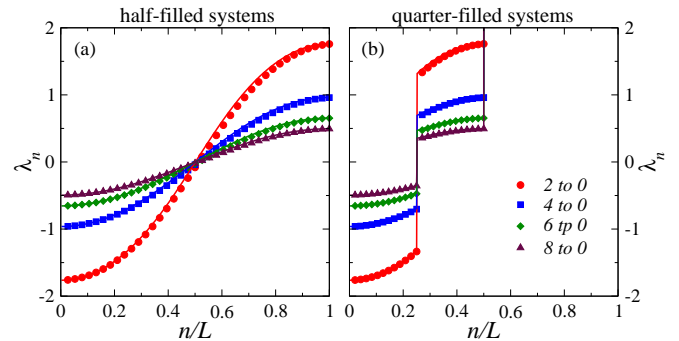}
\end{center}
\vspace{-0.7cm}
\caption{\label{fig:LagrangeMultipliers} (Color online) Lagrange multipliers in quenches from $A_I\neq0$ to $A_F=0$ 
for systems at half filling (a) and systems at quarter filling (b). (The results for quenches from 
$A_I=0$ to $A_F\neq0$ are the identical.) Symbols (lines) correspond to systems with 38 (380) sites in 
the half-filled systems [(a) and (b)] and to systems with 48 (480) sites in the quarter-filled systems. 
Once again, note that size effects for the Lagrange multipliers are negligible. 
Results are reported for quenches between $A_I=2$ and $A_F=0$, $A_I=4$ and $A_F=0$, $A_I=6$ and $A_F=0$, 
and $A_I=8$ and $A_F=0$. In the legend, we use the notation ``$A_I$ to $A_F$'' to label the plots.}
\end{figure}

In Fig.~\ref{fig:LagrangeMultipliers}, we show the values of the Lagrange multipliers for the same 
quenches and system sizes depicted in Fig.~\ref{fig:ConservedQuantities}. As expected from the expression 
for the Lagrange multipliers [Eq.~\eqref{eq:lagrangem}], they are a smooth function of the 
values of the conserved quantities (and exhibit negligible finite-size effects in Fig.~\ref{fig:LagrangeMultipliers}). 
One can then think of the conserved quantities, considered here to build the generalized ensembles,  
as additive in a coarse-grained sense. This follows if one realizes that by, increasing the system size, the 
Lagrange multipliers in a coarse-grained region do not change their values (Fig.~\ref{fig:LagrangeMultipliers}), 
but the sum of the expectation values of the conserved quantities in that region 
(Fig.~\ref{fig:ConservedQuantities}) grows proportionally to the increase of system size. Hence, 
effectively, the conserved quantities behave as additive. A discussion on the role of additivity 
of the conserved quantities in generalized ensembles can be found in Ref.~\cite{polkovnikov_sengupta_11}.

\section{Summary} \label{sec:summary}

We have studied the dependence on the initial state of the description of integrable systems 
after relaxation following a sudden quench. In general, integrable systems are not expected to
thermalize. Hence, we have focused on understanding special classes of initial states that 
lead to values of observables after relaxation that approach those in thermal equilibrium, when a 
control parameter is changed. One of our main findings is that, even for such initial states, 
thermalization does not occur as in nonintegrable systems. In the latter, the difference between 
the thermal expectation value of an observable and those after relaxation is expected to vanish 
in the thermodynamic limit. In the integrable systems discussed here, no matter the initial state 
selected (which is an eigenstate of another integrable system where the control parameter is 
one of the parameters of the initial Hamiltonian), the distribution of conserved quantities in 
the thermal ensembles differs from (but can be arbitrarily close to) that of the 
diagonal ensemble (or the GGE), and the difference does not vanish with increasing system size. 
Since the values of the conserved quantities constrain the outcome of the relaxation 
dynamics, the observables after relaxation do not reach thermal values in the thermodynamic 
limit. 

Another of our main findings is that what the control parameter is doing in those special
classes of initial states is tuning the distribution of conserved quantities to approach
thermal values. As a result, the initial states exhibit energy densities that are increasingly 
Gaussian like and entropies of their associated generalized ensembles that approach those of 
standard ensembles. Similarly to the behavior seen for the conserved quantities, the difference 
between the entropy per site in the generalized and standard ensembles remains nonzero in the 
thermodynamic limit. It can, however, be made arbitrarily small by changing the control parameter.
Interestingly, for the model considered here, the special initial states were found to be insulating 
ground states that approach products of single site wavefunctions. 

\vspace{-0.07cm}
\begin{acknowledgments}
This work was supported by the Office of Naval Research and the National Science Foundation under 
Grant No.~DMR-1004268. M.R. thanks A. Polkovnikov and L. F. Santos for useful discussions, and 
M. Fitzpatrick thanks J. Carrasquilla, E. Malatsetxebarria, and C. Varney for helpful suggestions.
\end{acknowledgments}

\end{document}